\documentclass[12pt,a4paper]{article}
\usepackage{amsmath,amsthm,amssymb,euscript,epsf,epsfig}
\usepackage{array}
\makeatletter
\renewcommand\section{\@startsection {section}{1}{\z@}%
                                   {-3.5ex \@plus -1ex \@minus -.2ex}
                                   {2.3ex \@plus.2ex}%
                                   {\normalfont\large\bfseries}}
\renewcommand\subsection{\@startsection{subsection}{2}{\z@}%
                                     {-3.25ex\@plus -1ex \@minus
                                     -.2ex}%
                                     {1.5ex \@plus .2ex}%
                                     {\normalfont\bfseries}}
\def\baselinestretch{1.2}
\newcommand{\be}{\begin{equation}}
\newcommand{\ee}{\end{equation}}
\newcommand{\bea}{\begin{eqnarray}\displaystyle}
\newcommand{\eea}{\end{eqnarray}}
\newcommand{\nn}{\nonumber}

\newcommand{\pd}{\partial}

\makeatletter
\@addtoreset{equation}{section}
\makeatother

\def\one{{\hbox{ 1\kern-.8mm l}}}
\def\zero{{\hbox{ 0\kern-1.5mm 0}}}

\def\cR{ {\cal{R}} }
\def\cRn{ {\cal{R}}_n }
\def\cRnp{ { { { \cal{R}}^+_n} }  }
\def\cRnm{ { {{\cal{R}}^-_n}}  }
\def\cRnpm{ { { { \cal{R}}^{ \pm }_n} }  }

\def\Prnp{ { \cal{P}}_{ {{\cal{R}}^+_n}}   }
\def\Prnm{ { \cal{P}}_{ {  {\cal{R}}^-_n} }      }
\def\Prn{  { \cal{P} }_{   {\cal{R}}_n    }   }

\begin{document}
{}~
{}~
hep-th/0603239\hfill\vbox{\hbox{QMUL-PH-05-15}
}\break

\vskip .6cm

\begin{center}
\Large \bf On Time-dependent Collapsing Branes\\
\Large \bf and Fuzzy Odd-dimensional Spheres
\end{center}

\medskip

\vspace*{4.0ex}

\centerline{\large C. Papageorgakis and  S. Ramgoolam ${}^{\dagger}$}

\vspace*{4.0ex}
\begin{center}
{\large Department of Physics\\
Queen Mary, University of London\\
Mile End Road\\
London E1 4NS UK\\
}
\end{center}

\vspace*{5.0ex}

\centerline{\bf Abstract} \bigskip
\noindent
We study the time-dependent dynamics of a collection
 of $N$ collapsing/expanding
$D0$-branes in type IIA String Theory.
We show that the fuzzy-$S^3$ and $S^5$ provide time-dependent
solutions to the Matrix Model of $D0$-branes and its DBI
generalisation. Some intriguing
cancellations in the calculation of the non-abelian DBI Matrix
actions result in the fuzzy-$S^3$ and $S^5$ having the same 
dynamics at large-$N$. For the Matrix model,  we
 find analytic solutions describing the time-dependent radius,
 in terms of Jacobi elliptic functions. 
 Investigation of the physical properties of these
configurations shows that there are no bounces for the trajectory of
the collapse at large-$N$. 
We also write down a set of useful identities  for fuzzy-$S^3$, 
fuzzy-$S^5$ and general fuzzy odd-spheres.

\thispagestyle{empty}

\vfill
\begin{flushright}
{\it ${}^{\dagger}${\{c.papageorgakis, s.ramgoolam\}@qmul.ac.uk}\\
}
\end{flushright}
\eject

\tableofcontents

\renewcommand{\baselinestretch}{1.05}  

\section{Introduction}
Fuzzy spheres of even dimensionality
\cite{madore,sphdiv,horam,kimhigh,kimura,kim2} have been 
at the centre of a number of phenomena in String and Matrix Theory \cite{bfss}, as well as the
study of the non-abelian DBI  action
\cite{coltuck,azbag,kabtay,chenlu1,chenlu2,clt,rst}. They 
enter classical solutions to $D0$-brane actions giving the 
microscopic description of  time-dependent $D0$-$D(2k)$ bound
states in type IIA, where the $D0$'s
expand into a fuzzy-$S^{2k}$ via a time-dependent analog of the Myers
 effect \cite{dielectric}.
 They  also describe a class of
static BIonic $D1$$\perp$$D(2k+1)$ brane intersections in type IIB, in which the
$D1$'s blow up into a funnel of fuzzy-$S^{2k}$ cross-section
\cite{gibb,malda,bion,nonabelian,koch,kochbhatta,bhattajam}. Both cases admit a 
macroscopic description in terms of the higher dimensional brane
worldvolume action, with
$N$ units of worldvolume  fluxes. The two pictures agree in the
large-$N$ limit. This agreement of the classical equations extends to
quadratic fluctuations for the
$D0$-$D2$ bound state  \cite{prt}. 
A full solution to the $1/N$ corrections for $S^2$, coming
from the implementation of the symmetrised trace ($STr$) prescription
for the non-abelian $D$-brane action,  has been given in
\cite{mac}.  The time-dependent and
static configurations for the fuzzy-$S^2$ are directly related via an $r\rightarrow 1/r$ duality
\cite{papram}. It is natural to consider extensions
 of these ideas to   systems involving fuzzy
spheres of odd  dimensionality.  Fuzzy odd-spheres
were constructed and studied  earlier in   \cite{sphdiv,zacram,geomodd}.
In \cite{sjab} the fuzzy 3-sphere algebra  was expressed as a 
quantisation of the Nambu bracket.  Subsequent work used 
the fuzzy 3-sphere in the context of $M2$$\perp$$M5$ intersections
 \cite{basuharv,copberg,no}.

This has provided a motivation to revisit fuzzy odd-spheres.
 In this paper, we study the fuzzy odd-sphere equations 
 in more detail and apply them to the 
 time-dependent process of $N$ $D0$'s blowing-up 
into a fuzzy-$S^3$ and $S^5$ respectively. Compared to the study of fuzzy
 even-spheres, these phenomena turn
 out to be significantly more involved.  Commutators of fuzzy
 odd-sphere
 matrices are not vanishing at large-$N$, hence 
calculating the symmetrised trace in that limit requires a 
non-trivial sum over orderings. 
After these sums are performed  we find the surprising result that
 the time evolution of the fuzzy-$S^3$ is identical to 
that of the fuzzy-$S^5$. 
 The rest of the paper is
 organised as follows. In section 2 we review the 
 fuzzy-$S^3$ and  higher-dimensional
 fuzzy odd-spheres. 
 A number of useful identities, which apply for odd-spheres of
 any dimensionality, are presented. Sections 3 and 4 focus on
 expressions for the
 particular cases of $S^3$ and $S^5$. Section 5 looks at the
 dynamics  of $N$ coincident $D0$-branes described by the Matrix DBI action. 
This is done by using an ansatz involving the fuzzy-sphere Matrices and 
a time dependent radius. This is inserted into the DBI action, to obtain 
a reduced action for the radius. It is shown, in Appendix C, that solutions 
to the reduced action also solve the Matrix equations of motion. 
 In section 6 we proceed 
to study the physical properties of these
 configurations, using a definition for the physical radius proposed
 in \cite{mac}, and find that there will be no bounces for
 large-$N$. The characteristic length scale of the system is
 $L=\sqrt{\pi}  \ell_s $ and independent of $N$. 
 In section 7 we show that both the fuzzy-$S^3$ and $S^5$ solve
 the  equations of motion in the Matrix Theory limit
 and yield solutions in terms of Jacobi elliptic
 functions. 
 In section 7  we
 discuss  a possible dual description of the fuzzy-$S^3$, in
 terms of a non-BPS $D3$-brane embedded in Euclidean space as
 a classical three-sphere.
 Finally, section 8 provides a summary and
 outlook. Appendix A deals with some of the  details on the
 evaluation of the $SO(4)$ $[X,X]$ term.  Appendix B discusses
 the non-associativity of the   projected $SO(2k)$ Matrix algebra,
 proposed to give a  non-associative 
deformation of the algebra of functions on 
$S^{2k-1}$ \cite{sphdiv}. We find, somewhat surprisingly, 
that the non-associativity does not vanish in the large-$n$ limit.
We describe an alternative product on the projected space of matrices 
which does become associative at large $n$.

\section{General  fuzzy odd-sphere equations with $SO(D)$ symmetry }
We start with a quick review of the construction of the fuzzy-$S^3$
and fuzzy-$S^5$ \cite{sphdiv,geomodd}. We are working with
matrices constructed by taking the symmetric $n$-fold tensor product of $V=V_+ \oplus
V_-$, where $V_+$ and $V_-$ are the two-dimensional spinor
representations of $SO(4)$, of respective positive and negative
chirality.  There are two projectors $P_\pm$, which project $V$ onto
$V_\pm$. In terms of the isomorphism $SO(4)=SU(2)\times SU(2)$ these
have respective spin $(2j_L, 2j_R)=(1,0)$ and  $(2j_L,
2j_R)=(0,1)$. The symmetrised tensor 
product space $Sym(V^{\otimes n})$, for every odd
integer $n$, contains a subspace $\mathcal R_n^+$ with
$(\frac{n+1}{2})$ factors of positive and $(\frac{n-1}{2})$ factors of
negative chirality. This is an irreducible representation of $SO(4)$
labelled by $(2j_L,2j_R)=(\frac{n+1}{2},\frac{n-1}{2})$. The projector
onto this subspace is in $End(Sym(V^{\otimes n}))$ and will be called $\mathcal
P_{\mathcal R_n^+}$. Equivalently, there is a subspace $\mathcal
R_n^-$ with spins $(2j_L,2j_R)=(\frac{n-1}{2},\frac{n+1}{2})$ and
projector $\mathcal P_{\mathcal R_n^-}$. The full space is then defined
  to be the direct  sum $\mathcal R_n=\mathcal R_n^+ \oplus \mathcal R_n^-$.
  The projector for this space is $\mathcal P_{\mathcal R_n}=\mathcal P_{\mathcal
      R_n^+} \oplus \mathcal P_{\mathcal R_n^-}$. The matrices $X_i$
  are in $End(\mathcal R_n)$
\be
X_i=\mathcal P_{\mathcal R_n}\sum_r \rho_r (\Gamma_i)\mathcal P_{\mathcal R_n}
\ee
 where $i=1,\ldots,4$, mapping $\mathcal R_{n^+}$ to $\mathcal R_{n^-}$ and vice versa. We
 can therefore re-express the above as a sum of matrices in
 $Hom(\mathcal R_n^+,\mathcal R_n^-)$ and $Hom(\mathcal R_n^-,\mathcal
 R_n^+)$
\be
X_i=\mathcal P_{\mathcal R_n^+}X_i\mathcal P_{\mathcal R_n^-} +
\mathcal P_{\mathcal R_n^-}X_i\mathcal P_{\mathcal R_n^+}
\ee
The product $X_i^2=C$ forms the quadratic Casimir of $SO(4)$. There is a
 set of generators   for the Matrix algebra
\bea\label{gens}
\nn  X_{i}^+ &=& \Prnm  ~ \sum_r \rho_r ( \Gamma_iP_+) ~  \Prnp    \\
\nn  X_{i}^{-} &=& \Prnp ~  \sum_r \rho_r ( \Gamma_i P_- ) ~ \Prnm   \\
\nn  X_{ij}^{+} &=&  \Prnp ~ \sum_r \rho_r \left(  \Gamma_{ij}P_+
\right)~ \Prnp \\
 \nn  Y_{ij}^{+}  &=& \Prnp ~ \sum_r \rho_r \left( \Gamma_{ij}  P_- \right)~  \Prnp\\
\nn 
  X_{ij}^{-} &=&  \Prnm  ~ \sum_r \rho_r \left(\Gamma_{ij} P_- \right) ~ 
\Prnm  \\
 Y_{ij}^{-}  &=& \Prnm ~  \sum_r \rho_r \left( \Gamma_{ij} P_+ \right)~    \Prnm 
\eea
where
\be
\Gamma_{ij}={ 1 \over 2 } [ \Gamma_i,  \Gamma_j ]
\ee
The coordinates of the sphere can be written as $X_i=X_i^+ +
X_i^-$ and one can also define the following combinations
\bea
\nn  X_{ij} &=& X_{ij}^{+} + X_{ij}^{-} \\
\nn  Y_{ij} &=&  Y_{ij}^{+} + Y_{ij}^{-} \\ 
\nn Y_i &=& X_i^+ - X_i^- \\ 
\nn \tilde X_{ij} &=&   X_{ij}^{+} -  X_{ij}^{-} \\
 \tilde Y_{ij} &=&  Y_{ij}^{+} - Y_{ij}^{-}
\eea
The generators above in fact form an over-complete set. It was observed 
\cite{nastase} that $X_i,Y_i$ suffice as a set of generators.  
 In the large-$n$ limit, the full Matrix algebra
 turns out to contain more degrees of freedom
 than the algebra of functions on the classical three-sphere.
 However, one can define an appropriate projection operation,
which then gives rise to the proper algebra of functions in the
large-$n$ limit. This projected Matrix algebra should be commutative and
associative\footnote{A discussion on the definition of this projection
and the large-$n$ behaviour of the associator can be
found in Appendix B.} at large-$n$.

For general fuzzy odd-dimensional spheres, $S^{2k-1}$, the Matrix
coordinates are matrices acting in a reducible representation
$\cRnp\oplus\cRnm$ of $SO(2k)$. The irreducible representations
$\cR_{\pm}$ have respective weights $\vec
r=(\frac{n}{2},\ldots,\frac{n}{2},\frac{\pm 1}{2})$, with $\vec r$ a
$k$-dimensional vector.  The matrices acting on the full space
$\cR=\cRnp\oplus\cRnm$ can be decomposed into four blocks
 $End(\cRnp)$, $End(\cRnm)$, $Hom(\cRnp,\cRnm)$ and
$Hom(\cRnm,\cRnp)$.

We will use the above to construct a number of useful
identities for any isometry group $SO(D)$, for $D=2k$ even.
There exist the following basic relationships \cite{sphdiv,zacram,geomodd} 
\bea\label{basics}  
( \Gamma_i \otimes \Gamma_i ) (P_+ \otimes P_+) &=&  0 \nn \\
 ( \Gamma_i \otimes \Gamma_i ) (P_- \otimes P_-) &=&  0 \nn \\
( \Gamma_i \otimes \Gamma_i ) (P_+ \otimes P_-) &=&  2 (  P_- \otimes P_+ ) 
 \nn \\
( \Gamma_i \otimes \Gamma_i ) (P_- \otimes P_+ )&=&  2 (  P_+ \otimes P_- )  
\eea 
For completeness, we give the explicit derivation. 
It is known from fuzzy even-spheres that $\sum_{\mu=1}^{2k+1} (
\Gamma_{\mu} \otimes \Gamma_{\mu}) $ 
acting on the irreducible subspace (which requires 
subtracting traces for $k>2 $) of $ Sym ( V \otimes V ) $ is equal to $1$. 
For any vector $v$ in this subspace, we have 
\bea 
 ( \Gamma_{\mu } \otimes \Gamma_{\mu }   ) v   = v 
\eea
Separating the sum over $\mu $ as $( \Gamma_{i} \otimes \Gamma_{i}) +
( \Gamma_{2k+1} \otimes \Gamma_{2k+1}) $, multiplying by $ (P_+ \otimes P_-) $ 
from the left, and using the Clifford algebra  relations 
proves the fourth equation above. The other equations are obtained similarly, 
by multiplying with an appropriate tensor product of projectors.     
>From these we derive
\bea\label{squaresD} 
X_i^2 &=& { ( n +1) ( n +D -1) \over 2 }  \equiv C \nn \\ 
X_{ij}X_{ij} &=& - { D  \over 4 }(n+1) (  n + 2D  -3 ) \nn \\ 
Y_{ij} Y_{ij} &=& -{ D \over 4 } ( n-1)(n + 2 D -5 ) \nn \\
X_{ij}Y_{ij} &=& { ( 4 - D ) \over 4 } ( n^2 -1 ) \nn \\      
X_iX_jX_iX_j &=&  ( 2 -D ) C \nn \\ 
\left[X_i, X_j\right] \left[X_j,X_i\right]  &=&  2 C (C+D-2)  
\eea 
and
\bea 
\left[ X_i,X_j\right] &=&  ( n+D-1) X_{ij} - X_{ijk} X_{k}  \nn\\
\left[ X_{j}, \left[ X_{j} , X_{i} \right] \right] &=&  2 (C+D-2) X_i  \nn\\
  X_{j} X_i X_j &=& ( 2 - D )  X_i \nn \\ 
 X_{ki}X_k &=& X_k X_{ik } = -  {  ( n + 2D -3 ) \over 2 }   X_i \nn \\ 
 Y_{ki}X_k &=&  X_k Y_{ik} =  { ( n -1 ) \over 2 }   X_i  \nn \\ 
 X_j X_k X_i X_j X_k &=& { 1 \over 4 }
 \bigl ( n^4 + 2n^3D + ( D^2 + 6 - 2D )n^2 \nn \\
&& + ( 6D - 2D^2 ) n - 3D^2 + 18 D
 - 23  \bigr ) X_i  
\eea
In the first equation of the second set we have used 
\bea 
 X_{ijk} = \Prnm \sum_{r}
 \rho_r ( \Gamma_{ijk} P_+ ) \Prnp +
 \Prnp \sum_{r} \rho_r ( \Gamma_{ijk} P_- ) \Prnm 
\nn 
\eea 
 where 
$ \Gamma_{ijk}$ is the normalised anti-symmetric product. 
It is useful to observe that 
\bea \label{commie}
\Prnm ~~~  \sum_{ r_1 \ne r_2 } ~~  \rho_{r_1}  ( \Gamma_k P_- ) ~~ 
\rho_{r_2} ( \Gamma_j P_+ )  ~~~ \Prnp = X_{k}^- X_{j}^+ + X_{jk}^{+} - 
{ ( n +1 ) \over 2 } \delta_{jk}
\eea 
It also follows that 
\bea 
&& \Prnm ~~~  \sum_{ r_1 \ne r_2 } ~~  \rho_{r_1}  ( \Gamma_k P_- ) ~~ 
\rho_{r_2} ( \Gamma_j P_+ )  ~~~ \Prnp  ~~~~+~~~~ ( + \leftrightarrow - ) 
\nn \\ 
&& = X_{k} X_{j} + X_{jk} - 
{ ( n +1 ) \over 2 } \delta_{jk} 
\eea 
where we have added the term obtained by switching the $+$ and $-$ from the 
first term. These formulae can be used to calculate
$ X_{i}X_{j}X_{k}X_i $ \footnote{Products of these will appear in the
  computation of determinants in the following sections.}
\bea 
X_{i}X_jX_k X_i  &&= X_{k}X_j \left[    { ( n-1) (n+D+1) \over 2 } + 2  \right] 
 - 2X_j X_k  \nn \\
&& ~~~ +  { ( n + 1 ) ( n+D-1) \over 2 } ( X_{jk} + Y_{jk} + \delta_{jk} )
 \nn \\ 
&& =  C ( -n , - D )~~ X_k X_j  + 2 [ X_k , X_j ] + 
C ( n , D ) ~~ ( X_{jk} + Y_{jk} + \delta_{jk} )
\eea 
In the last equality we have recognised that the coefficient of
 $ ( X_{jk} + Y_{jk} + \delta_{jk} )$ has turned out to be $ C = X_i^2 $ 
(\ref{squaresD}). We also made explicit the dependence of $C$ on $n, D $ 
writing $ C = C ( n , D ) $, and observed that the other numerical 
coefficient on the RHS is $ C ( -n , - D ) $.

In the large-$n$ limit there are significant simplifications to the 
above matrix identities
\bea\label{simplgN}  
 X_m X_i X_m &=& 0 \nn\\
 X_m X_i X_j X_m &=& C X_j X_i \nn \\ 
 A_{ij} A_{jk} &=& C ( X_i X_k + X_k X_i ) \nn \\ 
 X_{i} X_{p_1} X_{p_2 } \ldots X_{p_{2k+1} } X_i &=& 0 \nn \\ 
 X_{i}  X_{p_1} X_{p_2 } \ldots X_{p_{2k} } X_i &=& C  X_{p_2}\ldots X_{ p_{2k} } X_{p_1} \nn \\  
 \left[ X_i X_j , X_k X_l \right] &=&  0 \nn \\ 
 A_{kl} X_m &=& - X_{m} A_{kl} \nn \\ 
 X_i X_j X_k &=& X_{k} X_j X_i  
 \eea
 where, to avoid clutter, we have denoted
\be
A_{ij}=[X_i,X_j]
\ee 
and  $C\sim\frac{n^2}{2}$. From these it follows that
\bea\label{simpclgN}  
 X_m A_{ij} X_m &=& -C A_{ij} \nn \\ 
 \left[ A_{ij} , A_{kl } \right] &=& 0 \nn \\ 
 A_{ij} A_{ji} &=& 2 C^2\nn\\
 A_{ij}A_{jk}  A_{kl} A_{li} &=& 2 C^4 
\eea 
As an example of how these simplifications occur,
consider the last equality of (\ref{simplgN}). 
As explained at the beginning of this section, we can decompose a 
string of operators such  as $ X_i X_j X_k = X_i^+ X_j^- X_k^+ +  X_i^- X_j^+ X_k^- $.  
Writing out $X_i^+ X_j^- X_k^+ $ 
\bea 
 X_i^+ X_j^- X_k^+ &=& \Prnm 
 \sum_{ r_1, r_2 , r_3 } \rho_{r_1} ( \Gamma_i P_+ )
\rho_{r_2} ( \Gamma_j P_- ) \rho_{ r_3 } ( \Gamma_k P_+ )\Prnp \nn \\ 
& =& \Prnm \sum_{ r_1 \ne  r_2  \ne  r_3 } \rho_{r_1} ( \Gamma_i P_+ ) 
\rho_{r_2} ( \Gamma_j P_- ) \rho_{ r_3 } ( \Gamma_k P_+ )  \Prnp\nn \\ 
& =& \Prnm \sum_{ r_1 \ne  r_2  \ne  r_3 }  \rho_{ r_3 } ( \Gamma_k P_+ )
\rho_{r_2} ( \Gamma_j P_- )\rho_{r_1} ( \Gamma_i P_+ )  \Prnp  \nn \\ 
& =& X_k^+ X_j^- X_i^+
\eea 
In the second line we used the fact that the terms 
with coincident $r$'s, such as $ r_1 =r_2 $, are sub-leading 
in the large-$n$ limit.  
There are $ { \cal { O } } ( n^3 )  $ terms of type
 $ r_1 \ne r_2 \ne r_3 $ while there are  $ { \cal { O } } ( n^2 )  $
terms of type  $ r_1 = r_2 \ne r_3 $ and ${\cal O} (n) $ terms of type 
$ r_1 =r_2 = r_3 $. 
 In the third line, we used the fact that   
operators acting on non-coincident tensor factors commute. 
We find 
\bea 
 X_i X_j X_k &=& X_i^+ X_j^- X_k^+ +  X_i^- X_j^+ X_k^- \nn \\  
            &=& X_k^+ X_j^- X_i^+ +  X_k^- X_j^+ X_i^-  \nn \\ 
            &=& X_k X_j X_i 
\eea
Similar manipulations along with the basic relationships (\ref{basics})
 lead to the rest of the  formulae in (\ref{simplgN}).

\section{On the equations for  fuzzy-$S^3$ } 
Specialising to the case of the fuzzy-$S^3$ we can deduct further
Matrix identities. Squaring the generators
\bea
X_i^2 &=& { ( n+1)(n+3) \over 2 } \nn \\ 
X_{ij}^2 &=&  - ( n +1) ( n+5)  \nn \\ 
Y_{ij}^2 &=& - ( n -1 ) ( n +3)  \nn \\ 
X_{ij}Y_{ij} &=& 0 \nn
\eea 
Note that $X_{ij} Y_{ij} =0 $ in the case of $SO(4)$.
This product is not zero for general 
$D$. We also have
\bea \label{eqsfrs3} 
[ X_i , X_j ] &=& { ( n + 3 ) \over 2 } X_{ij} - { ( n +1 ) \over 2 }  Y_{ij} 
\nn\\
X_{j}X_{ki}X_j &=&  { (n+1)(n+5)\over 2 }  Y_{ki} \nn \\
X_j Y_{ki} X_j &=& { (n-1)(n+3) \over 2 } X_{ki} \nn \\  
X_{ki } X_k &=& -  {  n +5 \over 2 } X_i \nn \\
Y_{ki} X_k &=&  { n - 1 \over 2 } X_i \nn \\ 
X_k X_j X_k &=& -2 X_j \nn \\
X_{j}X_{ki}X_jX_k &=& {(n-1)( n+1)(n+5)\over 4 } X_i \nn \\ 
X_{j}Y_{ki}X_jX_k &=& - {(n-1)( n+3)(n+5)\over 4 } X_i \nn \\ 
X_jX_kX_iX_jX_k &=& { 1 \over 4 }   ( n^2 + 4n -1 )^2  X_i 
\eea  
In the second pair of equations of the above set, 
note that we might have expected $ X_{j}X_{ki}X_j $ 
to be a linear combination of $X_{ki}$ and $ Y_{ki} $ 
but only $ Y_{ki} $ appears. This follows directly from the
transformation properties of these operators under $SO(4)$.

We can  compute $ X_jX_kX_iX_jX_k$ directly and get an answer which works for any 
$D$. Alternatively we can make use of the $S^3$ identities  
\bea 
X_jX_kX_iX_jX_k  &=& X_j ( [ X_k , X_i ] + X_i X_k ) X_j X_k  \nn \\ 
    &=& { (n+3) \over 2 } X_j X_{ki}X_jX_k - { ( n+1) \over 2 }X_jY_{ki} X_j X_k + X_j X_i X_k X_j X_k
\eea
Using the formulae (\ref{eqsfrs3}) we see that
the contributions from the first two terms are equal.
The two computations of this object of course agree. 

It is worth
noting here that the decomposition of the commutator $[ X_i, X_j ]$
into a sum over $X_{ij}$ and $Y_{ij}$ should be expected.  In
\cite{sphdiv} a complete $SO(4)$ covariant basis of matrices acting on
$\cRn$ was given in terms of operators corresponding to self-dual
and anti-self dual Young diagrams. According to that analysis, the most
general anti-symmetric tensor with two free indices should be a linear combination
of the following structure
\be
\sum_r \rho_r (\Gamma_{ij} )
\ee
with any allowed combination of $P_\pm$ on $\cRn^{\pm}$, 
where the $SO(4)$ indices on the $\Gamma$'s have been suppressed  for
simplicity. Note that the coefficients multiplying the above basis
elements include contractions with the appropriate $\delta$ and $\epsilon$-tensors.
For $SO(4)$ the antisymmetric two-index tensors are
(anti)self-dual and $\epsilon_{ijkl}\Gamma_{kl}P_{\pm}= \pm 2
\Gamma_{ij}P_{\pm}$, with $\Gamma_5 P_\pm=\mp \Gamma_1\ldots
\Gamma_4 P_\pm$.
 Contractions with $\delta$ are of course ruled out
for symmetry reasons. As a consequence, everything can be expressed in terms
of $X_{ij}$, $Y_{ij}$. The same procedure can be used to show that
every composite object with one free $SO(4)$ index $i$ can be reduced to
be proportional to $X_i^{\pm}$. The allowed linearly
independent basis elements are
\bea
\nn && \sum_r \rho_r(\Gamma_i)\\
\nn && \sum_{r\neq s} \rho_r(\Gamma_{ij})\rho_s(\Gamma_k)\delta_{jk}
\sim \sum_s \rho_s(\Gamma_i)\\
&& \sum_{r\neq
  s}\rho_s(\Gamma_{jk})\rho_s(\Gamma_l)\epsilon_{ijkl}\sim\sum_s \rho_s(\Gamma_i)
\eea
It is easy to see explicitly that the last two quantities are proportional to
$X_i$, when evaluated on $\cRnpm$. Since we should be able
to express any object with one free index in terms of this basis, it will
necessarily be proportional to $X_i$.

\section{On the equations for  fuzzy-$S^5$ }
For the fuzzy-$S^5$ we only present a few specific identities
that will appear in the following sections. The commutator decomposes into
\be
[X_i , X_j ] = (n+5) X_{ij}- X_{ijk} X_k
\ee
Alternatively we can  express this as
\be
X_i^\mp X_j^\pm-X_j^\mp X_i^\pm = (n+1) X_{ij}^\pm+\Prn\frac{i}{6} \epsilon_{ijklmn} \left[\sum_{r\neq s}
\rho_r(\Gamma_{lmn}) \rho_s(\Gamma_k)\right] X_7\Prn
\ee
where 
$\Gamma_7 P_\pm=\pm i\Gamma_1\ldots\Gamma_6 P_\pm$ and $X_7\Prn=\Prnp-\Prnm$.
There is no expression
for $ [ X_{i} , X_{j} ] $ as a linear combination
of only $ X_{ij} $ and $ Y_{ij} $, unlike the case of
the fuzzy-$S^3$. This is not surprising since the $SO(6)$
covariant basis for two-index antisymmetric tensors will now include
terms of the form
\bea
\nn && \sum_r \rho_r (\Gamma_{ij})\\
\nn && \sum_r \rho_r (\Gamma_{klmn})\epsilon_{ijklmn} \sim \sum_s
\rho_s(\Gamma_{ij})\\
&& \nn \sum_{r\neq s}
\rho_r(\Gamma_{kl})\rho_s(\Gamma_{mn})\epsilon_{ijklmn}\sim \sum_s
\rho_s(\Gamma_{ij})\\
&& \sum_{r\neq s}\rho_r(\Gamma_{klm})\rho_s(\Gamma_n)\epsilon_{ijklmn}
\eea
Note that the last expression is not proportional to
$\sum_r\rho_r(\Gamma\Gamma)$.
 We can once again show that any composite
tensor with one free $SO(6)$ index $i$ should be proportional to
$X_i^\pm$, just as in the $SO(4)$ case. We have
\bea
\nn &&\sum_r \rho_r(\Gamma_i)\\
\nn &&\sum_{r\neq s}
\rho_r(\Gamma_{ij})\rho_s(\Gamma_{k})\delta_{jk}\sim  \sum_s \rho_s(\Gamma_i)\\
\nn && \sum_{r\neq s}
\rho_r(\Gamma_{jkl})\rho_s(\Gamma_{lm})\epsilon_{ijklmn}\sim\sum_s \rho_s(\Gamma_i) \\
&& \sum_{r\neq s\neq t}
\rho_r(\Gamma_{jk})\rho_s(\Gamma_{kl})\rho_t(\Gamma_m)\epsilon_{ijklmn}
\sim\sum_t \rho_t(\Gamma_i)
\eea
as can be easily verified for any $P_\pm$ combination on $\cRnpm$.

\section{The Fuzzy-$S^{2k-1}$  matrices and  DBI with symmetrised trace
 }
In this section we will substitute the ansatz 
\be\label{ansatz}
\Phi_i=\hat{R}(\sigma,t )X_i
\ee
into the Matrix DBI action of $D1$-branes to obtain an effective 
action for $ \hat R $. We will show in Appendix C that solutions to  the 
reduced equation of motion also give solutions to the Matrix DBI
 equations of motion.  To
begin with, we will give the most general expressions for time-dependent
$D$-strings. Dropping the dependence on the 
spatial direction $\sigma$  will reduce the
problem to that of time-dependent $D0$-branes. Assuming a static
ansatz will lead to $D1$-brane fuzzy funnels.

\subsection{Fuzzy-$S^3$}
The low energy effective action for $N$ $D$-strings with no
worldvolume gauge
field and in a flat background is given by the non-Abelian
Dirac-Born-Infeld action \cite{dielectric}
\be\label{sd1}
S=-T_1\int d^2\sigma STr\sqrt{-\det\left[
\begin{array}{cc}
\eta_{ab} &\lambda\partial_a\Phi_j \\
-\lambda\partial_b\Phi_i &Q_{ij}
\end{array}
\right]}\equiv -T_1\int d^2\sigma STr\sqrt{-\det (M)}\quad\
\ee
where
\be
 Q_{ij} = \delta_{ij}+i\lambda\big[\Phi_i ,\Phi_j\big]
\ee
and $\lambda =2\pi \ell_s^2$. The determinant can be explicitly
calculated keeping in mind the symmetrisation procedure. The result is
\bea\label{dets3}
\nn -\det (M) &=& 1+{\lambda^2\over 2}\Phi_{ij}\Phi_{ji}
+\lambda^4 \left({1 \over 8}(\Phi_{ij}\Phi_{ji})^2-{1\over 4}
\Phi_{ij}\Phi_{jk}\Phi_{kl}\Phi_{li} \right)\\
 & &\,\,\, +\lambda^2\partial^a\Phi_i\partial_a\Phi_i+
\lambda^4\left({\partial^a\Phi_k\partial_a\Phi_k
\Phi_{ij}\Phi_{ji}\over 2}-\partial^a\Phi_i\Phi_{ij}\Phi_{jk}
\partial_a\Phi_k\right)
\eea
Considering  the ansatz (\ref{ansatz}) with 
 $i=1,\ldots ,4$ describes the fuzzy-$S^3$.
The $X_i$'s are $N$$\times$$N$ matrices of
$SO(4)$, as defined in section 2. Their size is given by
$N=\frac{1}{2}(n+1) (n+3)$. Substituting into (\ref{dets3}) we get
\bea\label{strings3}
\nn -\det (M) & = &1+{\lambda^2\over 2}\hat{R}^4 A_{ij}A_{ji}
+{\lambda^4\over 4}\hat{R}^8 \left({(A_{ij}A_{ji})^2 \over 2}-A_{ij}A_{jk}A_{kl}A_{li}\right) \\
\nn & & +\lambda^2(\partial^a\hat{R})(\partial_a\hat{R}) X_i X_i+
\lambda^4 \hat{R}^4(\partial^a\hat{R})(\partial_a\hat{R}) \left( {X_k X_k
A_{ij}A_{ji}\over 2}-X_i A_{ij}A_{jk}X_k\right)\\
\eea

At this point we need to implement the symmetrisation of the
trace. In order to simplify the problem, this procedure can be carried
out in two steps. We first symmetrise the terms that lie under the
 square root. We then perform a binomial expansion and symmetrise
 again. The even-sphere  cases that were considered in \cite{papram}
 didn't involve this complication, since the commutators 
$[X_i,X_j]$, $[A_{ij},A_{kl}]$ and $[X_i,A_{jk}]$ turned out to be 
 sub-leading in $n$. Thus, for large-$N$ the square root argument was already
 symmetric and gave  a simple result straight away. Here, however, the
 $[X_i,X_j]$ and $[X_i,A_{jk}]$ yield a  leading-$n$ contribution and the
 symmetrisation
 needs to be considered explicitly. 

 From now on, we will focus
 completely on the  time-dependent problem of $N$ type IIA $D0$-branes and drop
 the $\sigma$-direction. Then the  ansatz (\ref{ansatz}) will be
 describing a dynamical effect of collapsing/expanding branes. 
 Had we chosen to consider the
 static version of the above action, we would have a collection of
 coincident $D$-strings blowing-up into a funnel of higher 
 dimensional matter with an $S^{2k-1}$ cross-section.

\subsubsection{Vanishing symmetrised trace contributions } 
 The terms involving only  $A$'s are already symmetric, since the
 commutator of commutators, $[A_{ij},A_{kl}]$,
is sub-leading in the large-$N$ limit. 
>From (\ref{simplgN}) and (\ref{simpclgN}) it follows immediately that the coefficient of 
$ \hat R^8 $ in (\ref{strings3}) vanishes. The latter can be expressed 
as 
\be 
Sym ( A + B ) = 0 
\ee 
where $ A = \frac{  A_{ij} A_{ji} A_{kl}A_{lk} } { 2 } \equiv A_1A_2A_3A_4 $, 
$ B = A_{ij} A_{jk} A_{kl} A_{li }\equiv B_1 B_2 B_3 B_4 $. In this case,
  as we have already mentioned,
 the $A_{ij}$'s commute and  $ Sym ( B ) = B , Sym( A)  = A $.

When we expand the square root we obtain terms 
of the form 
$ Sym ( C  ( A + B )^k ) $ where $C $ is a product of operators 
$ C_1C_2 \ldots  $, e.g. $ C = ( X_iX_i )^{n}$.
 It is easy to see that these also vanish. 
The symmetrised expression will contain terms 
of the form 
\be\label{sigonCAB}  
\sum_{ l=0}^{k } { k \choose  l } \sum_{ \sigma} 
\sigma ( C_1 \ldots C_n  ( A_1 \ldots A_4 )^l  ( B_1 \ldots B_4 )^{k-l}  ) 
\ee 
The sum over $\sigma $ will contain terms where the
$l  $  copies of  $A_1 \ldots A_4$ and the $ k-l $ copies of  $ B_1 \ldots B_4$ are
 permuted amongst each other and also amongst the $C$ factors. 
Due to the relations in (\ref{simplgN}) and (\ref{simpclgN}), we can permute 
the $4l$ elements from $ A^l$ through the $C$'s and $B$'s 
to collect them back in the form of $A^l$. Likewise for $B^{k-l}$. 
Since $A_{ij}$ elements commute with other $A_{kl}$ and 
anti-commute with $X_k $, we will pick up, in this re-arrangement, 
 a factor of $ (-1) $ raised to the number of times 
 an $A_{ij}$ type factor crosses an $X_k$ factor. This is a factor 
that depends on the permutation $ \sigma $ and on $k$ but not 
on $l$, since the number of $A_{ij}$'s coming from $A$ and $B$ do not depend 
on $l$. We call this factor $ N ( \sigma , k ) $. The above sum takes the form 
\be\label{nosig}  
\sum_{ \sigma}  N ( \sigma , k )  ( C_1 \ldots C_n)
\sum_{ l=0}^{k } { k \choose  l }     ( A_1 \ldots
A_4 )^l  
( B_1 \ldots B_4 )^{k-l}   
\ee 
This contains the expansion of $ (A + B  )^k  $ with no permutations 
that could mix the  $A, B$ factors. Therefore it is zero.

Similarly, we can show that
the coefficient of $ { \hat R }^2  \dot {\hat R}^2 $ 
in the determinant (\ref{strings3})
is zero. This requires a small calculation of summing over $24$
permutations
\footnote{Or $6$ if we fix one element using cyclicity.}.  
The relevant formulae are 
\bea\label{steps} 
 X_i X_i A_{jk} A_{kj} &=& \phantom{-}2 C^3 \nn \\ 
 X_i A_{jk} X_i A_{kj} &=& - 2C^3 \nn \\ 
 X_i X_j A_{ik} A_{kj} &=& \phantom{-}C^3 \nn \\ 
 X_i A_{ik} X_j A_{kj} &=& -C^3
\eea 
which follow from (\ref{simplgN}). The outcome is again $ Sym ( A + B ) =  0  $, where 
$ A = \frac{ X_k X_k A_{ij}A_{ji}}{2 } \equiv  A_1 A_2 A_3 A_4   $ and
$ B = X_k X_i A_{ij}A_{jk} \equiv B_1 B_2 B_3 B_4 $. In this case, 
we do not have   $ Sym ( A ) = A , Sym ( B ) = B $, since the factors 
within  $A,B$ do not commute. We can repeat the above
arguments to check $Sym(A+B)^k$. Start with
 a sum of the form 
\be 
\sum_{l=0}^k { k \choose l } \sum_{ \sigma}
\sigma (  A^k B^{k-l} ) 
\ee 
Because of the permutation in the sum, there will be terms where 
 $ X_{i} X_{i} A_{jk}A_{kj} $ has extra $X$'s interspersed in between. 
Such a term  can be re-collected 
into the original  form at the cost of introducing a sign factor, 
a factor $E(n_i)  = ( 1 + (-1)^{n_i} )/2  $, where $n_i$ is the number of $X$'s
 separating the two $X_i$, and also introducing a permutation of the
 remaining $X$'s.  We will describe this process in more detail, 
but the important fact is that these factors will be the same 
when we are re-collecting $ X_{i}X_{k} A_{ij} A_{jk}$, i.e. the index
structure doesn't affect the combinatorics of the re-shuffling.
The first step is to move the $A_{jk}A_{kj} $ all the way to the right, 
thus picking up a sign  for the number of $X$'s one moves through
during the process. We then have 
\bea\label{reshfst}  
&& X_{i}~ (  X_{p_1} ... X_{p_m} ) ~  X_i ~ ( X_{q_1} .. X_{q_n} ) ~ 
 A_{jk}A_{kj} \nn \\ 
&& = E ( m )  ( X_{p_2} ... X_{p_m} X_{p_1} )~  X_i X_i ~ 
 ( X_{q_1} ... X_{q_n} ) ~  A_{jk}A_{kj} \nn \\ 
&&= E(m)   ( X_{p_2} ... X_{p_m} X_{p_1} )  ( X_{q_1} ... X_{q_n} )
 X_iX_i A_{jk}A_{kj}  
\eea
In the last line $E(m) = (1 + (-1)^m)/2 $ is $1$ if $m$ is even 
and zero otherwise. 
If instead we are considering  $ X_{i}X_{k}A_{ij} A_{jk}$ we have 
\bea\label{reshsec}  
&&  X_{i}~ (  X_{p_1} ... X_{p_m} ) ~  X_k ~ ( X_{q_1} ... X_{q_n} ) ~ 
 A_{ij}A_{jk} \nn \\ 
&&= X_i ~ (  X_{p_1} ... X_{p_m} ) ~ X_k ~  ( A_{ij}A_{jk} ) ~ 
( X_{q_1} ... X_{q_n} ) \nn \\ 
&&= X_i ~ (  X_{p_1} ... X_{p_m} ) ~ X_k ~  ( C (X_iX_k + X_kX_i)  ) ~ 
( X_{q_1} ... X_{q_n} ) \nn \\ 
&&=  X_i ~ (  X_{p_1} ... X_{p_m} ) ~ C^2 X_i ~ 
( X_{q_1} ... X_{q_n} ) \nn \\ 
&&= E ( m) C^3  ( X_{p_2} ... X_{p_m} X_{p_1} )( X_{q_1} ... X_{q_n} ) \nn \\ 
&&= E(m ) ( X_{p_2} ... X_{p_m} X_{p_1} )  ( X_{q_1} ... X_{q_n} )
 X_iX_k A_{ij}A_{jk} 
\eea 
The last lines of (\ref{reshfst}) and (\ref{reshsec}) 
 show that the rules for re-collecting $ A $ and $B$ from 
more complicated expressions, where their components 
have been separated by a permutation, are the same. 
Hence, 
\be 
\sum_{l=0}^k {k \choose l}\sum_{\sigma} \sigma ( A^l B^{k-l}) 
\ee 
can be re-written, taking advantage of the fact that 
terms with different values of $l$ only differ in substituting 
$A$ with $B$. This does not affect the combinatorics of re-collecting. 
Finally, we get
\be 
\sum_{\sigma} F ( \sigma ,k ) \sum_{l=0}^k { k \choose l }  A^l B^{k-l}   
= 0 
\ee 
The $F(\sigma,k)$  is obtained from collecting all the sign 
and $E$-factors that appeared in the discussion of the above re-shuffling. 

When other operators, such as some generic $C$,  are involved 
\be 
\sum_{l=0}^k  { k \choose l } \sum_{\sigma}  \sigma( C  A^l B^{k-l}  )   
\ee 
the same argument shows that
\be 
 \sum_{\sigma}  { \tilde \sigma}   ( C ) 
 \sum_{l=0}^k { k \choose l } A^l B^{k-l}    = 0 
\ee 
Note that $ \sigma $ has been replaced by $ \tilde \sigma $ because the 
process of re-collecting the powers of $ A,B $,
as in (\ref{reshsec}) and (\ref{reshfst}),  involve a re-shuffling 
of the remaining operators. The proof of $ Sym ( A + B ) =0  $ can also be presented 
along the lines of  the above argument. Then
\bea 
 X_{i} A_{jk} X_{i}  A_{kj} &=& - X_iX_i A_{jk} A_{kj} \nn \\ 
 X_{i} A_{ij} X_{k} A_{jk}   &=& - X_{i}  X_{k} A_{ij} A_{jk}
\eea
which, combined with $X_iX_i A_{jk} A_{jk} = 2  X_{i}  X_{k} A_{ij} A_{jk} $, 
gives the vanishing result.

\subsubsection{Non-vanishing symmetrised trace contributions } 
After the discussion in the last section, we are only left to consider
\be\label{only}
 STr\sqrt{-\det(M)}=\sum_{m=0}^\infty STr\left(\frac{\lambda^2}{ 2}\hat{R}^4 
A_{ij}A_{ji}+\lambda^2(\partial^a\hat{R})(\partial_a\hat{R}) X_i X_i\right)^m
\binom{1/2}{m}
\ee
We derive the following formulae for symmetrised traces
\bea\label{symmtrs}  
{ STr ( XX )^m \over N }  &=&  C^m { (m!)^2 2^m \over (2m)! } \nn \\
{ STr ( ( XX)^{m_1} (AA)^{m_2} ) \over N } &=&  2^{m_1+m_2} C^{m_1 + 2m_2} 
 { ( m_1 + m_2 )!\over m_2!}  
{m_1!  (2m_2)! \over ( 2m_1 + 2m_2 ) ! }  \nn \\   
{ STr  ( AA )^m \over N }  &=&  2^{m}C^{2m} 
\eea  
To calculate the first line, note that we have
to sum over all possible permutations of $ X_{i_1}X_{i_1} X_{i_2}X_{i_2} \ldots X_{i_m}
X_{i_m}$. For all terms where the two $ X_{i_1}$'s are separated by an even 
number of other $X$'s we can replace the pair by $C$. Whenever the 
two are separated by an odd number of $X$'s they give a sub-leading 
contribution and therefore can be set to zero in the large-$N$ expansion. 
In doing the averaging we treat the two $i_1$'s as distinct objects and 
sum over $(2m)!$ permutations. For any ordering let us label the positions 
from $1 $ to $2m$.  To get a non-zero answer, we need one set 
$ i_1 ... i_m $ to be distributed amongst $m$ even  places in $m!$ ways 
and another set of the same objects to be distributed amongst the odd positions 
in $m!$ ways. There is a factor $2^m$ for permutations of the two copies 
for each index. As a result we obtain the  $C^m { (m!)^2 2^m \over (2m)! } $.

Now consider the second line. By cyclicity we can always fix the 
first element to  be an $X$. There are then $ ( 2m_1 + 2m_2 -1 ) ! $ 
permutations of the $ (2m_1 -1)$ $X$-factors. As we have seen, 
in the large-$N$ limit  $ A X = - X A $. Reading towards 
the right, starting from the first $X$, suppose we have $p_1$ 
$A$'s followed by an $X$, then $p_2$ $A$'s followed by an $X$, etc. 
This is weighted by $(-1)^{p_1 + p_3 + \cdots + p_{2m_1-1}} $. 
Therefore, we sum over all partitions of $2m_2$, including a multiplicity 
for different orderings of the integers in the partition, and weighted 
by the above factor. This can be done by a mathematical package such
as Maple in a variety of cases and gives 
\be 
{ ( m_1 + m_2 -1)! \over m_2! ( m_1 -1 )! } { ( 2m_2 )! ( 2 m_1 -1 )! \over 
 ( 2m_1 + 2m_2 -1 )! } 
\ee
The denominator $  ( 2m_1 + 2m_2 -1 )! $ comes from the number of permutations
which keep one $X$ fixed. The above can be re-written 
in a way symmetric under the exchange of $ m_1$ with $m_2 $ 
\be 
\frac{ ( m_1 + m_2 )!} {m_1!  m_2! } {  (2m_1 )! ( 2 m_2)! \over 
 ( 2m_1 + 2m_2 )! } 
\ee 
The factor $ \frac{ (m_1!)^2 2^{m_1}} {(2m_1)! } $ comes 
from the sum over permutations of the $X$'s. 

We describe another way to derive this result. This time we will not  
use cyclicity to fix the first element in the permutations 
of $ (XX)^{m_1} ( AA)^{m_2} $ 
 to be $X$. Let there be $p_1$ $A$'s on the left, then 
one $X$, and $p_2$ $A$'s  followed by another $X$ and 
so on, until the last $X$ is followed by 
$p_{2m_1+1} $  $X$'s. We will evaluate this string
by moving all the $A$'s to the left, picking up 
a sign factor $ (-1)^{ p_2 + p_4  + \ldots + p_{2m_1} }$ in the process. 
This leads to a sum  over $p_1\ldots p_{2m_1 +1}$ which can be re-arranged 
by defining $ P  =  p_2 + p_4 + \ldots + p_{2m_1} $. 
$P $ ranges from $ 0$ to  $ 2m_2$ and is the total number 
of $A$'s in the even slots. For each fixed $P$ there is a 
combinatoric factor of 
$ \tilde C ( P , m_1 ) =  { ( P + m_1 - 1 )! \over P! m_1! }$ 
from arranging the $P$ objects into  $m_1$ slots.  
There is also   a factor  $ \tilde C (2m_2 -  P , m_1 +1  )$
from arranging the remaining $(2m_2 -P)$ $A$'s into the $m_1 +1 $ 
positions. These considerations lead to 
\bea 
&&\sum_{p_{2m_1}=0}^{2m_2}  \sum_{ p_{2m_1 - 1} =0 }^{ 2m_2 - p_{2m_1}}
 \cdots \sum_{p_1=0}^{ 2m_2 -  p_2 - \cdots - p_{2m_1} }
( -1)^{ p_2 + p_4 + \cdots + p_{2m_1  }  }  \nn\\
&&= \sum_{P=0 }^{2m_2} (-1)^{P} \tilde C ( P , m_1  ) 
\tilde C ( 2m_2 - P , m_1 +1  ) \nn \\ 
&&= { (m_1 + m_2 )! \over m_1! m_2! } \nn 
\eea 
The factor obtained above is multiplied by 
$ (2m_2)! $ since 
all permutations among the $A$'s give the same answer.
Summing over permutations of $X$'s give the 
extra factor $ 2^{m_1} (m_1!)^2 $. Finally, there is a  
normalising denominator of $(2m_1 + 2m_2 )!$.   Collecting these 
and the appropriate power of $C$  gives 
\be 
2^{m_2} C^{m_1 + 2m_2 } 
{ (m_1 + m_2 )! \over m_1! m_2! } 
 {2^{m_1} (m_1!)^2 ( 2m_2) ! \over ( 2m_1 + 2m_2)! } 
\ee 
which agrees with the second line of 
(\ref{symmtrs}). 

\subsection{Fuzzy-$S^5$}
We will now turn to the case of the fuzzy-$S^5$. The starting action
will be the same as in (\ref{sd1}).  However, the ansatz 
incorporates six non-trivial transverse scalars $\Phi_i=\hat R(\sigma, t ) X_i$,
where $i=1,\ldots,6$. The $X_i$'s are $N$$\times$$N $ matrices of
$SO(6)$, as defined in section 2, with their size given by
$N=\frac{1}{192}(n+1)(n+3)^2 (n+5)$. By truncating the problem to the purely
time-dependent configuration, this system represents a dynamical
process of $N$ $D0$-branes expanding into a fuzzy-$S^5$ and then
collapsing towards a point. The static truncation provides an 
analogue of the static fuzzy-$S^3$ funnel, with a collection of $N$ $D$-strings
blowing-up into a funnel with a fuzzy-$S^5$ cross-section. 

The calculation of the
determinant  yields the following result
\bea\label{dets5}
\nn -\det (M) &=& 1+{\lambda^2\over 2}\Phi_{ij}\Phi_{ji}
+\lambda^4 \left( {1\over 8}(\Phi_{ij}\Phi_{ji})^2-{1\over 4}
\Phi_{ij}\Phi_{jk}\Phi_{kl}\Phi_{li} \right)\\
\nn & &\,\,\, +\lambda^6\left({(\Phi_{ij}\Phi_{ji})^3\over 48}\right.
-{\Phi_{mn}\Phi_{nm}\Phi_{ij}\Phi_{jk}\Phi_{kl}\Phi_{li}\over 8}
+\left.{\Phi_{ij}\Phi_{jk}\Phi_{kl}\Phi_{lm}\Phi_{mn}\Phi_{ni}\over 6}
\right)\cr\\
\nn & &\,\,\, +\lambda^2\partial^a\Phi_i\partial_a\Phi_i+
\lambda^4\left({\partial^a\Phi_k\partial_a\Phi_k
\Phi_{ij}\Phi_{ji}\over 2}-\partial^a\Phi_i\Phi_{ij}\Phi_{jk}
\partial_a\Phi_k\right) \\
\nn & &\,\,\, -\lambda^6\left(
{\partial^a\Phi_m\partial_a\Phi_m
\Phi_{ij}\Phi_{jk}\Phi_{kl}\Phi_{li}\over 4}
-{\partial^a\Phi_i\partial_a\Phi_i
(\Phi_{jk}\Phi_{kj})^2
\over 8}\right. \\
 & &\,\,\, +\left. {\partial^a\Phi_i
\Phi_{ij}\Phi_{jk}\partial_a\Phi_k
\Phi_{ml}\Phi_{lm}\over 2} -\partial^a\Phi_i
\Phi_{ij}\Phi_{jk}\Phi_{kl}\Phi_{lm}
\partial_a\Phi_m\right) 
\eea
Once again we will need to implement the symmetrisation procedure,
just as we did for the fuzzy-$S^3$.
 The structure of the terms within the square root for $D=6$ is
 almost the same as for $D=4$. The only difference is that we will now
 need to include expressions of the type $Sym(A+B+C)$ and $Sym(A+B+C+D)$,
 coming from the two new $\cal{O}$$(\lambda^6 )$ terms. Consider the first
 of these for example. After
 expanding the square root, we will need the multinomial series
 expansion
\be
\sum_{n_1,n_2,n_3\ge 0}
\frac{n!}{n_1!n_2!n_3!}A^{n_1}B^{n_2}C^{n_3}\qquad \textrm{with}\qquad n_1+n_2+n_3=n
\ee
and analogously for the expression with four terms. The initial
multinomial coefficient will separate out and  we will then have to deal with
the permutations, just as we did for the binomial terms. Note that the symmetrisation  
 discussion in the former section did not make any use of the fact that $D=4$. All the
 simplifications that occurred by taking the large-$N$ limit
 and the  combinatoric factors, which came from the re-shuffling
 of the operators, were derived for representations of
 $SO(2k)$ with $k$ not specified. It is straightforward to see that the former arguments
 will carry through to this case. The effect of the permutations for
 any possible combination of terms will simply introduce a common
  pre-factor, multiplied by the multinomial coefficient of
 the original term.  One can easily verify that, with the help of the
 large-$N$ matrix identities from section 2,  all the  terms multiplying powers
of $\lambda$ higher than $2$ in (\ref{dets5}) will give a zero
contribution. Therefore, the substitution of the ansatz will give
\be
\nn STr\sqrt{-\det(M)}=\sum_{m=0}^\infty STr\left(\frac{\lambda^2}{ 2}\hat{R}^4 
A_{ij}A_{ji}+\lambda^2(\partial^a\hat{R})(\partial_a\hat{R}) X_i X_i\right)^m
\binom{1/2}{m}
\ee
This is, somewhat surprisingly, exactly what appeared in (\ref{only}). 
It is intriguing that there is such a universal description for both the $D=4$
and $D=6$ problems.

\section {The large-$N$ dynamics of fuzzy odd-spheres. }
The discussion of the previous section allows us to write 
the Lagrangian governing the collapse/expansion of the fuzzy 3-sphere 
as well as the fuzzy 5-sphere. 
\bea
\nn \mathcal L &=& - \sum_{m=0}^\infty \sum_{k=0}^m {1/2\choose
  m}{m\choose k} (-1)^{m-k}\left(\frac{\lambda^2 \hat R^4}{2}
\right)^k (\lambda^2 \dot{\hat R}^2 )^{m-k}\; STr\left[(A_{ij}A_{ji})^k(X_i X_i)^{m-k} \right] \\
&=& - \sum_{m=0}^\infty \sum_{k=0}^m {1/2\choose
  m}{m\choose k}
(-1)^{m-k}s^{2(m-k)}r^{4k}\frac{m!}{(2m)!}\frac{(2k)!}
{k!}2^{m-k}(m-k)!
\eea
where we have defined 
\bea\label{sumvars}
r^4 &=& { \lambda^2  }   {\hat R }^4 C^2 \nn\\ 
s^2 &=& { \lambda^2 } { \dot {\hat R }}^2 C 
\eea
 Alternatively we can express this as two infinite sums with $n+k=m$
\be\label{lags3}
\mathcal L = - \sum_{n=0}^\infty \sum_{k=0}^\infty{1/2\choose
  n+k}{n+k\choose k}
(-1)^{n}s^{2n}r^{4k}\frac{(n+k)!}{(2(n+k))!}\frac{(2k)!}
{k!}2^{n}n!
\ee
>From this we can calculate the energy of the configuration to get
\be
E= - \sum_{n=0}^\infty \sum_{k=0}^\infty (2n-1) {1/2\choose
  n+k}{n+k\choose k}
(-1)^{n}s^{2n}r^{4k}\frac{(n+k)!}{(2(n+k))!}\frac{(2k)!}
{k!}2^{n}n!
\ee
The first sum can be done explicitly  and gives
\be
E= \sum_{n=0}^\infty
 \left(\frac{s^2}{2}\right)^n \;  _2 F_1
\left(\frac{1}{2},n-\frac{1}{2},n+\frac{1}{2}; -r^4\right)
\ee
There is an identity for the $_2F_1$ function 
\be
\nn _2F_1(a,b,c;z)=\; (1-z)^{-a}\; _2F_1\left(a,c-b,c;\frac{z}{z-1}\right)\;,\qquad\textrm{for}
\qquad z \notin (1,\infty)
\ee
We can use  this to re-express the energy sum as
\be
E=\sum_{n=0}^\infty
\left(\frac{s^2}{2}\right)^n \frac{1}{\sqrt{1+r^4}}\;  _2 F_1
\left(\frac{1}{2},1,n + \frac{1}{2}; \frac{r^4}{1+r^4}\right)
\ee
We need one more step to complete the
evaluation. The integral representation for the hypergeometric
function is
\be
\nn _2F_1(a,b,c;z)=\frac{\Gamma(c)}{\Gamma(b)\Gamma(c-b)}\int_0^1
\rho^{b-1}(1-\rho)^{-b+c-1}(1-\rho z)^{-a} d\rho
\ee
for $Re(c)>Re(b)>0$ and $|Arg(1-z)|<\pi$. These conditions are
satisfied for $n\neq 0$ so we will split the sum into two parts. The
$n=0$ part simplifies to just $\sqrt{1+r^4}$, while the rest is
\bea
\nn &&\sum_{n=1}^\infty
\left(\frac{s^2}{2}\right)^n \frac{1}{\sqrt{1+r^4}}\;  _2 F_1
\left(\frac{1}{2},1,n+\frac{1}{2}; \frac{r^4}{1+r^4}\right) \\
 &=&\sum_{n=1}^\infty
\left(\frac{s^2}{2}\right)^n \frac{1}{\sqrt{1+r^4}}\left(n-\frac{1}{2}\right)
\int_0^1 \frac{(1-\rho)^{n-\frac{3}{2}}}{\sqrt{\left(1-\rho \frac{r^4}{1+r^4}\right)}}d\rho
\eea
By first summing over $n$ and then performing the integration over $\rho$
we get a result,
which when added to the $n=0$ piece gives the full answer for the energy
\bea\label{energy}
E &=&\frac{\sqrt{1+r^4}\sqrt{s^2/2+r^4}\left(r^4(s^2/2-1)-s^2/2\right)
+r^4 (s^2/2-1)(s^2/2)
 \tanh^{-1}{\left(\sqrt{\frac{r^4+s^2/2}{r^4+1}}\right)}}
{(r^4+s^2/2)^{3/2}(s^2/2-1)}\nn \\ 
&=& \sqrt { 1 + r^4 }  \left( 1 - { s^4 \over (s^2-2) ( s^2 + 2r^4 ) } \right)
 + { r^4 s^2 \over  2 ( r^4 +s^2/2)^{3/2}  } 
\tanh^{-1}{\left(\sqrt{\frac{r^4+s^2/2}{r^4+1}}\right)} 
\eea 
We can use the same method to obtain the explicit form of the
Lagrangian (\ref{lags3}), restoring all the appropriate  dimensionful parameters
\be\label{lagrangian}
S=-\frac{N}{g_s \ell_s}\int dt
\left(\sqrt{1+r^4}-\frac{s^2}{2\sqrt{r^4+s^2/2}}
\tanh^{-1}{\frac{\sqrt{r^4+s^2/2}}{\sqrt{1+r^4}}} \right)
\ee
We can perform expansions of the above expression for small values of
the $r$ and $s$ variables
\bea 
E ( r  \sim 0 , s ) &=&   { 1 \over ( 1 -s^2/2 ) }
     + r^4 { { 1 \over s^2 -2 } + { \sqrt{2} r^4 \over s } 
  \tanh^{-1}{ s \over \sqrt{2}}   } + \cdots           
  \nn \\ 
E ( r , s  \sim 0  ) &=&    \sqrt{ 1 + r^4 } + { s^2 \over {2 r^2 }} 
\tanh^{-1}{\left(\sqrt{\frac{r^4}{r^4+1}}\right)} + \cdots 
\eea 
Also for large $r$ and small $s$ 
\bea 
E ( r,s )&=&  \left( r^2 + { 1\over 2 r^2 } - { 1 \over 8  r^6 } \right) + 
            \left( { 1 \over 8 r^6 } + { \ln( 2r^2 )  \over 2 r^2 } \right) s^2 
         \nn \\ &&    + \left(   { 3 \over 16 r^6 } - { 3  \ln ( 2 r^2
	   ) \over 8 r^6 } + 
{ 3 \over 8 r^2 } \right) s^4 
             + \left(  { 5 \over 32 r^2 } -  {5 \over 32 r^6 } \right)
	     s^6 + \cdots
\eea        
One of the features of the fuzzy even-spheres, namely the fact that
they admit $r=t$ type solutions is also true here. 
This statement translates into having  an $s^2/2=1$
solution to the $\pd_t E=0$ equations of motion. It is easy to check
that this holds. It is unfortunate that the energy formula includes
an inverse hyperbolic tangent of $(r,s)$. This leads to a
transcendental relationship between the two variables and  prevents us from conducting
an analysis similar to \cite{papram}, which would give the time of
collapse, a possible solution to the radial  profile and the
configuration's periodicities.

\subsection{Constancy of the speed of light }\label{constancy}
In order to explore  the physical properties of our configurations, we
 will use a new  definition of the physical radius for any $S^{2k-1}$ fuzzy sphere
 \bea 
\nn   R_{phys}  ^2 &=& \lambda^2 \lim_{m\rightarrow\infty}
\frac{STr (\Phi_i \Phi_i)^{m+1}}{STr (\Phi_i \Phi_i)^{m}}\\
 &=& \lambda^2 \hat R ^2
\lim_{m \rightarrow \infty }\frac{
   STr ( X_i X_i )^{m+1}}  { STr ( X_i X_i )^{m}} \nn \\
&=& { \lambda^2 C { \hat R }^2 \over 2 } 
\eea 
where we have evaluated the matrix products in the large-$N$ limit.
This will guarantee that the series defining the Lagrangian 
 will converge for $ \dot R_{phys } = 1 $. 
The definition can be interpreted as an application of the 
principle of constancy of the speed of light. It was introduced in
 \cite{mac} where it also gave the correct results for the fuzzy even-sphere
 problems at both large and finite-$N$. We would like to highlight
 that the  above is not the same thing as requiring local Lorentz invariance.
This is because the form of the summed series is not the 
same as in special relativity. For example the action takes the form 
$  \int  { dt \over { 1 - { \dot R_{phys}}^2  } } $
 at $ R_{phys} =0$, rather than the standard
$ \int   dt \sqrt{{ 1 - { \dot R_{phys}}^2  } } $ which appears 
in the large-$N$ limit of the $D0$-$D(2k)$ systems.
 Modifications of the standard relativistic 
form arise in the  study of fuzzy even-spheres at finite-$N$ \cite{mac}.

In terms of the $(r,s)$ variables, which appear in the 
expressions for the energy in the previous section, we have
\bea\label{scls}  
r &=& { \sqrt {2} R_{phys}  \over \sqrt{\lambda} }=\frac{R_{phys}}{L_{\textrm{odd}}}  \nn \\ 
s &=& \sqrt{2}  \dot R_{phys} 
\eea 
The physical singularity at  $ \dot R_{phys}
=1$ corresponds to $s^2=2$. For later
convenience, we will further define the normalised
dimensionless variables $\sqrt{2}(\hat r, \hat s)= (r,s)$. This
implies that $\hat s=1$ is the speed of light.   
The characteristic   length scale of the system for fuzzy 
odd spheres, appearing in (\ref{scls}), is
$L_{\textrm{odd}}=\sqrt{\frac{\lambda}{2}}$. This should be contrasted 
with $ L_{\textrm{even}} = \sqrt{ N \lambda \over 2 }$.  
 In the fuzzy-$S^2$ problem 
we were able to keep $L$ finite when $\ell_s \rightarrow 0 $ while
$\sqrt N \rightarrow \infty$. 
In the present case we cannot do so. If we take 
$\ell_s\rightarrow 0$ we lose the physics of the fuzzy odd-spheres. 
This is compatible with the idea that they should be  related to 
some tachyonic configuration on an unstable higher dimensional dual
brane. These tachyonic modes become 
infinitely massive as $ \ell_s \rightarrow 0 $.

\subsection{Derivatives, no-bounce results, accelerations} 
We can use the above results to get a picture of the
nature of the collapse/expansion for the $D0$'s blown-up into a
fuzzy-$S^{2k-1}$.
 We can explore whether there will be a bounce in the
trajectory, as was done in \cite{rst,mac} for the finite-$N$ dynamics
of fuzzy even-spheres, by looking for zeros of constant
energy contour plots in $(r,s)$. This can be simply seen by a zero of
the first derivative of the energy with respect to $s$, for constant $r$.
For our case this  is
\bea 
 { \partial E \over \partial \hat s  }|_{\hat r} &=& 
{  2 \sqrt { ( 1 + 4 \hat r^4) } \hat s^3
 (  \hat s^2 + \hat r^4 (10 - 6\hat s^2) ) \over (1-\hat s^2)^2 (4\hat
  r^4 + \hat s^2 )^2 } 
\nn \\ 
 && + {  4\hat r^4 ~\hat s ~ ( 8\hat r^4 -\hat s^2 ) 
\tanh^{-1} ( { \sqrt { 4 \hat r^4 + \hat s^2 } \over \sqrt { 1 +4 \hat
    r^4 } } ) \over
 (4\hat r^4 + \hat s^2 )^{5/2} } 
\eea 
We have checked numerically that the above expression is 
non-zero for $0<\hat s<1$ and $\hat r>0$. Hence there will be no bounce
and the configuration will classically collapse all the way to zero
radius. In our treatment so far, we have assumed that higher
derivative $\alpha '$ corrections can be neglected. This
statement translates into requiring that higher
commutators should be small. This condition gives 
$[ \ell_s\Phi_i,[\ell_s\Phi_i,~ ]] \ll 1$
and, with the use of  $[X_i,[X_i,]]=n^2$ for large-$n$, it implies
 that $\hat r \ll \sqrt{ \pi \over 2 }  \sim 1 $. This corresponds to 
\be\label{upbnd}  
R_{phys} < \ell_s 
\ee
The other relevant quantity in investigating the validity of
the action is the proper acceleration, which should be small. 
 This is defined as
\be
\alpha =\gamma^3 \frac{d^2 \hat r}{d\tau^2}=\gamma^3 \hat s
\frac{d\hat s}{d\hat r}=-\gamma^3
\frac{\partial_{\hat r} E}{\partial_{\hat s} E}
\ee
with $\gamma = (1-\hat s^2)^{-1/2}$. The derivative of the energy with
respect to $\hat r$ is
\bea 
  { \partial E \over \partial {\hat r}  }|_{\hat s} &=& 
{ 8 \hat r^3(\hat s^4 +16\hat r^8 (\hat s^2 -1)+4\hat r^4 \hat s^2(4
  \hat s^2 -3) )  \over \sqrt{ 1 + 4\hat r^4 } (\hat s^2-1 ) ( 4\hat r^4 +
  \hat s^2 )^{2 } }\nn \\
&& + \frac{16 \hat r^3 \hat s^2 (\hat s^2 - 2 \hat
  r^4) \tanh^{-1}({\sqrt{\frac{4 \hat r^4+ \hat s^2}{1 + 4 \hat
	r^4}}})}{(4 \hat r^4 + \hat s ^2)^{5/2}}
\eea
>From the above we get a complicated expression for the proper
acceleration in $\hat r$ and $\hat s$. Since the energy relation, combined
 with the boundary condition that the velocity at $\hat r_0$ is zero, 
gives a transcendental equation for $\hat r$ and $\hat s$, we can't predict
the behaviour of the velocity for the duration of the collapse. 
However, for small $\hat r$ the proper acceleration simplifies to
\be
\alpha = \frac{4 \hat r^3 \left(\hat s +2 (\hat s^2
  -1)\tanh^{-1}({\hat s})\right)}{\hat s\sqrt{1-\hat s^2}} + {\cal O}(\hat r^5)
\ee
In the small $\hat r$ limit, we can see that the velocity $\hat s$ will also be
small throughout the trajectory\footnote{If we take $r^4 \ll 1$ in
  (\ref{energy}) we find that $s\sim 0$.} and we will be within the Matrix
Theory limit, which we describe in the next section. For these values
the proper acceleration is small enough and the action is valid
throughout the collapse. If we were to give the configuration some
large  initial velocity, we have numerical evidence that there will be
 a region where the proper acceleration is small enough for the action
 to be valid but could change sign. We leave the further investigation
 of the relativistic regime for the future.

\section{The Matrix Theory (Yang-Mills) limit } 
It is interesting to consider the Matrix Theory limit of the action
for both the $S^3$ and the $S^5$  cases. Consider equation
(\ref{only})
\be
\nn S = -\frac{1}{g_s \ell_s} \int dt STr \sqrt{1+\frac{\lambda^2}{2}\hat R^4
  A_{ij}A_{ji}- \lambda^2\dot{\hat R}^2 X_i X_i}
\ee
For small $\hat R$ and  small $\dot{\hat R}$, the above yields
\be
S=- \frac{N}{g_s \ell_s}\int dt \left( 1+ \frac{\lambda^2}{4}\hat R^4
A_{ij}A_{ji}-\frac{\lambda^2}{2}\dot{\hat R}^2 X_i X_i \right)
\ee
The equations of motion for both $S^3$ and $S^5$
will be in dimensionless variables
\be
\ddot r = -2 r^3
\ee
and will be solved exactly by radial profiles of the form
\be
r(\tau)= r_0 ~ Cn\left( \sqrt{2}~ r_0~\tau,\frac{1}{\sqrt{2}} \right) 
\ee
where $r_0$ is a parameter indicating the value of the initial radius
at the beginning of the collapse. However, we need to prove that
having a solution to the above reduced action is equivalent to solving
the Matrix equations of motion. Starting from equation (\ref{sd1}) and
doing a small $\lambda$ expansion, we arrive at
\be\label{ym}
S= -\frac{1}{g_s\ell_s}\int dt Tr \left(1+
\frac{\lambda^2}{4}[\Phi_i,\Phi_j][\Phi_j,\Phi_i]
-\frac{\lambda^2}{2}\partial_t \Phi_i \partial_t \Phi_i \right) 
\ee
The equations of motion are
\be\label{ymphi}
\frac{\partial^2 \Phi_i}{\partial t ^2} = -[\Phi_j,[\Phi_j,\Phi_i]]
\ee
and upon substituting the ansatz, $\Phi_i = \hat R(t) X_i $, and using
the matrix  identities for any $D$ we get
\be
\ddot{\hat R} = - \hat R^3 (C+D-2)
\ee
It is easy to verify that, by directly substituting the ansatz into
(\ref{ym}) and then calculating the equations of motion, we will get
the same result. Therefore, any solution of the reduced action
will also be a solution of the full Matrix action for any $N$.

The time of collapse
can be calculated by using conservation
of energy,  in the same way as for $S^2$ \cite{rst,papram}. Expressed in
terms of the characteristic length-scale of the theory,  the answer will be
$T\sim L^2/R_0$. However, as we have already noted, the parameter $L$
in the odd-sphere case does not contain a factor of
$\sqrt{N}$. As a consequence, the collapse of fuzzy-$S^3$ and $S^5$
spheres at large-$N$ and for $R_0\ll L$, will occur much faster than
for the fuzzy-$S^2$. It will also not be experiencing any 
finite-$N$ modifications of the kind observed in \cite{mac}.

\section{Towards a dual 3-brane picture for fuzzy-$S^3$  }
In the spirit of the dualities that have been established between
higher and lower dimensional branes for the case of even spheres
\cite{kim2,rst,bion,nonabelian,koch}, it
is reasonable to expect that there should be a dual description for
the system of $D0$-branes blowing-up into a fuzzy-$S^3$. The $D0$'s
 couple to the $RR$ three-form potential, via a term proportional to  $ \int dt ~STr(
C^{(3)}_{tij}[\Phi_j,\Phi_i]+C^{(3)}_{kij}~\dot{\Phi}_k[\Phi_j,\Phi_i])$
in the Chern-Simons part of the action. They also couple to
to the five-form potential via $\int dt~STr(
 C^{(5)}_{tijkl}[\Phi_i,\Phi_j][\Phi_k,\Phi_l])$, although by 
 expanding the commutator terms and checking that
 the leading order term\footnote{That is the
one with no coincidences.} in $\epsilon_{ijkl} X_i X_j X_k X_l$ 
evaluates to zero, one sees that this is
sub-leading in $n$. The overall
trace will ensure that the net higher brane charge will be zero. Therefore, at large-$n$
we will only see an effective multi-pole coupling
to the $D2$-brane charge. 

The simplest candidate for a dual construction is a collection of coincident
$D3$'s embedded in Euclidean space as a classical 3-sphere, which in Type IIA are non-BPS. As a
consequence, we expect to have an effective field theory on the
compact unstable  $D3$ ($UD3$)
worldvolume, which will incorporate a real tachyon field and some $SU(M)$
non-trivial worldvolume flux, where $M$ is the number of $UD3$'s. 
 The most general  description will be in terms of the non-abelian, tachyonic
 $Dp$-brane action proposed in \cite{garousi}.
In the case at hand, there are no non-trivial transverse scalars and
we are working in a flat background, 
in a spherical embedding. The DBI part of the $UD3$-brane action will then reduce to
\bea
S&=&-\frac{\mu_3}{g_s}\int d^{4}\sigma ~STr\left(V(T)
\sqrt{-\det(P[G]_{ab}
+\lambda F_{ab}+\lambda D_a T D_b T}) \right)\nn\\
&=&-\frac{\mu_3}{g_s}\int d^{4}\sigma ~STr\left(V(T) \sqrt{-\det(M)} \right)
\eea
The symmetrisation procedure should be implemented amongst all non-abelian expressions
of the form $F_{ab}$ and $D_aT$ and on $T$ of
 the potential, which  is well
approximated  by $V(T)\sim e^{- T^2} $. The gauge field strength 
and covariant derivative of tachyons are
\bea
F_{ab}&=&\pd_aA_b-\pd_bA_a-i[A_a,A_b]\nn\\
D_aT&=&\pd_aT-i[A_a,T]
\eea
It is straightforward to calculate the determinant.
 The result, in spherical coordinates $(\alpha_1,\alpha_2,\alpha_3)$, is
\bea
\nn\sqrt{-\det(M)}&=&\sqrt{g}\left[ (1-\dot{R}^2-\lambda
  D_0 T D_0 T)\left( R^6 -\frac{\lambda^2
    R^2}{2}F_{ij}F^{ji}\right) - R^4 \lambda^2
F_{0i}F_0^{\phantom{0}i}\right.\\
\nn &&+(1-\dot{R}^2)\left(\lambda R^4 D_i T D^i T
  -\lambda^3 \left(\frac{D_i T D^i T F_{jk}F^{kj}}{2}-D^i T F_{ij} F^{jk} D_k T \right)\right)\\
 \nn&&\left.-2 \lambda^3 R^2 D_0 T D_j T F_{0i} F^{ij} +\frac{\lambda^4}{4}\left(
 \frac{F_{0i}F_0^{\phantom{i}i}F_{jk}
   F^{kj}}{2}-F_{0i}F^{ij}F_{jk}F^k_{\phantom{k}0}\right)
\right]^{1/2}\\
\eea
where $g=\sin^4 {\alpha_1} \sin^2 {\alpha_2} $ is the determinant of
the unit three-sphere metric.

The Chern-Simons part of the non-BPS brane action which is  of interest to us, has
been discussed in \cite{kennwilk,kraus} and is given in terms of the
curvature of the super-connection
\bea
S_{CS} &\sim &  \mu_3\int C\wedge STr ~ e^{\frac{\mathcal F}{2\pi}}\nn \\
&\sim &  \mu_3\int C\wedge STr ~ e^{\frac{1}{2\pi}( F
  -T^2+DT)}
\nn\\
 &\sim &  \frac{\mu_3}{8\pi^2}\int C^{(1)}\wedge STr ~ \left(( F\wedge DT )~ W(T)\right)
\eea
where only odd-forms are kept in the exponential expansion, the
$DT\wedge DT\wedge DT$ term vanishes because of the overall
symmetrisation and $W(T)=e^{-\frac{T^2}{2 \pi}}$. 
There will also be a term proportional to $\int C^{(3)}\wedge
STr(DT)$. If we want to match the two descriptions, we should require that the
overall $D0$-charge is conserved. We therefore impose that
\be\label{sphsym}
\frac{1}{8\pi^2}STr~\left( (F \wedge DT) W(T) \right)=\frac{N}{\textrm{Vol}_{S^3}} \Omega_3
\ee
 where $  \Omega_3 $  is the $SO(4)$ invariant volume form on the 3-sphere with angles
 $(\alpha_1, \alpha_2, \alpha_3)$ and  $N$ is an
integer.  By restricting to  the
worldvolume of a single brane, i.e. having a $U(1)$ gauge symmetry for
the gauge and tachyon fields, this condition translates in components into the following expression
\be
 F_{[ij}\partial_{k]} T ~ W(T) =  4N  \epsilon_{ijk}
\ee
where $\epsilon_{\alpha_1 \alpha_2 \alpha_3}=\sqrt{g}$. 
Contraction with $\epsilon^{ijk}$ results in
\be\label{2nd}
\epsilon^{ijk} \partial_iT F_{kj} =  -\frac{24 N}{W(T)}
\ee
while contraction with $\partial^iT F^{kj}$ and then use of
the above equation gives
\be\label{detsim}
\frac{\partial_i T \partial^i T F_{jk}F^{kj}}{2}-\partial^i T F_{ij} F^{jk} \partial_k T
=-\frac{144 N^2}{W(T)^2}
\ee
Using this last expression we can simplify the
$UD3$ action just by  $SO(4)$ symmetry of the charge conservation
condition. We obtain
\bea\label{ud3}
\nn S &=&-\frac{\mu_3}{g_s }\int d\sigma^4\; \sqrt{g}\;
V(T)\;  \left[ (1-\dot{R}^2-\lambda
  \dot{T}^2)\left( R^6 -\frac{\lambda^2
    R^2}{2}F_{ij}F^{ji}\right) \right.\\
\nn &&-2 \lambda^3 R^2 \dot{T} \partial_j T F_{0i} F^{ij}- R^4 \lambda^2
F_{0i}F_0^{\phantom{0}i}+(1-\dot{R}^2)\left(\lambda R^4 \partial_i T \partial^i T
  +\frac{144 \lambda^3 N^2}{ W(T)^2} \right)\\
 &&\left. +\frac{\lambda^4}{4}\left(
 \frac{F_{0i}F_0^{\phantom{i}i}F_{jk}
   F^{kj}}{2}-F_{0i}F^{ij}F_{jk}F^k_{\phantom{k}0}\right)
\right]^{1/2}
\eea
 The equations of motion for this configuration are involved.
Nevertheless, note that all the individual  terms in the above action should be
 scalars of $SO(4)$. This means that we should have both  
$\partial_i T \partial^i T$ and $T^2$  independent of the
 angles and a constant $T$ cannot satisfy the spherical
 symmetry condition (\ref{sphsym}). 
 Hence, there is no worldvolume theory for a compact unstable brane system in
three dimensions with $SO(4)$ symmetry and we should be looking at
least at higher numbers of coincident non-BPS branes in order to find a dual
description.

We may expect the matching between the $D0$ and the higher brane pictures to be more tricky than
 in the even-sphere case. In the even-sphere,
 the upper bound on the validity of the lower brane description
increases with $N$, and allows an overlap 
with the higher brane description \cite{bion}. As can be seen from 
 (\ref{upbnd}), the upper bound does not increase with $N$ for the 
fuzzy odd-sphere case. The separation of the degrees of freedom 
of the fuzzy-$S^3$ Matrix algebra \cite{geomodd} into geometrical and internal 
 also suggests 
that a simple relation to a non-abelian theory is not possible. 
However, this does not preclude a relation via a non-trivial 
renormalisation group flow, analogous to that proposed 
in \cite{basuharv} in the context of the application of the fuzzy 
three-sphere to the $M2$$\perp$$M5$ intersection. 

\section{Summary and Outlook}
In this paper we provided a set of formulae for
general fuzzy odd-spheres and studied them as solutions of 
 Matrix DBI $D0$-brane actions and their Matrix Theory (Yang-Mills)  
limit. After implementing the symmetrised trace, which in the fuzzy-odd 
sphere case requires a non-trivial sum over orderings even at large-$N$, 
we found the same equations of motion for the fuzzy-$S^3$ and $S^5$.
  We proved that solutions to the reduced
DBI action also solve  the full Matrix  equations of motion. For the
Matrix Theory  limit, we gave exact
expressions for these solutions in terms of Jacobi elliptic
functions. The study of the physical properties of these
systems showed that the classical collapse will proceed all the way to
the origin. 

 Given that we have now established the fuzzy-$S^3$  (and $S^5$) 
 as solutions to stringy Matrix Models, we can study the action 
for fluctuations. Using the remarks in \cite{geomodd} on the 
 geometrical structure of the Matrix algebras, we expect that it should 
 be possible to write the action in terms of fields on a higher 
dimensional geometry: $S^2 \times S^2$  for the $S^3$ case and 
 $SO(6)/ ( U(2) \times U(1) ) $ in the case of $S^5$. 
 It will be intriguing 
to clarify the geometry and symmetries of this action.   

A very interesting open problem is to 
identify a macroscopic large-$N$ dual description of the
fuzzy-$S^3$ and $S^5$ systems that we have described. The main difficulty 
lies in constructing 
$SO(2k)$-invariant finite energy time-dependent 
solutions describing tachyons coupled to gauge fields on odd-dimensional 
spheres. 
 The spherical symmetry restrictions that
we discussed here should  facilitate the task of
investigating a  non-abelian solution. 
A dual description, and agreement with the microscopic picture at the
level of the action and/or equations of motion,  would not
 only provide us with a new check of the
current effective worldvolume actions for non-BPS branes, but also
give the possibility of constructing cosmological toy models
of bouncing universes with three spatial dimensions.
The gravitational back-reaction of such time-dependent spherical brane
 bound states would also be of interest as a possible avenue 
towards physically interesting time-dependent versions of the AdS/CFT duality.  
 A non-trivial
extension  to the problem would be the addition of angular
momentum. This could provide a stabilisation mechanism
for fuzzy odd-spheres in the absence of the right $RR$ fluxes, which provide
a simple stabilisation mechanism for fuzzy even-spheres. The study of
 finite-$N$ effects and the embedding of these
systems in more general backgrounds \cite{js1,js2}  provide other possible
avenues for future research. Finally it would be interesting to get 
a better understanding of the relation between the fuzzy-odd
sphere constructions in this paper and those 
of  \cite{JLR1,JLR2} involving fibrations over projective spaces.

\bigskip
{ \bf Acknowledgements}:
We would like to thank David Berman, Neil Copland, John Ward and
Kostas Zoubos for
useful discussions. The work of SR is supported by
a PPARC Advanced Fellowship. CP would like to acknowledge a
QMUL Research Studentship. This work was in
part supported by the EC Marie Curie Research Training Network
MRTN-CT-2004-512194.

\begin{appendix}
\section{Derivation of $ [ X_{i} , X_j ] $ for fuzzy-$S^3$ }
Here we include part of the calculation that led to the
commutator decomposition for the fuzzy-$S^3$, using two possible ways
of evaluating $[X_i,X_j]^2$. 
One is a direct method and starts by writing out the
commutators
\bea 
\nn [X_i,X_j][X_j,X_i] &=& 2 X_i X_j X_i X_j -2 X_i X_j X_j X_i\\
 &=& 2 X_i^+ X_j^- X_i^+ X_j^- + 2 X_i^- X_j^+ X_i^- X_j^+ - 2C^2
\eea
A straightforward calculation for any $D$ gives
\be
X_i^\mp X_j^\pm X_i^\mp X_j^\pm = -\frac{(D-2)(n+1)(n+D-1)}{2}
\ee
The two combine into 
\be
[X_i, X_j][X_j,X_i]=-\frac{(n+1)(N+D-1)}{2}\left[2D-4+(n+1)(n+D-1) \right]
\ee
 The other is based on the
 decomposition of any antisymmetric tensor of $SO(4)$ with two free
 indices in terms of the appropriate  basis elements 
\be
[X_i, X_j] = \alpha ~ X_{ij}+ \beta ~Y_{ij}
\ee
where $\alpha$ and $\beta$ are some yet undetermined
coefficients. Then
\be
[X_i,X_j][X_j,X_i]= \alpha^2 ~X_{ij}X_{ji}+ \beta^2 ~Y_{ij}Y_{ji}
\ee
since for $D=4$ we have seen that $X_{ij}Y_{ij}=0$. Deriving the
identities for $X_{ij}X_{ji}$ and $Y_{ij}Y_{ji}$ is a simple task.
Using these results, it is easy to evaluate the above expression and
compare against what we get from the
straightforward calculation. The outcome is
\be
\alpha= \frac{(n+3)}{2}  \qquad\textrm{and}\qquad \beta =-\frac{(n+1)}{2}
\ee
This result can also be checked against calculations of $ X_{ij} [X_{i} , X_j ] $
and $ Y_{ij}  [X_{i} , X_j ] $.

\section{The large-$n$ limit of the fuzzy-$S^3$ projected algebra}

In \cite{geomodd} it was proposed that there is a simple prescription for
obtaining the space of functions on $S^{2k-1}$ by performing a
projection of the Matrix algebra onto symmetric and traceless
representations of $SO(2k)$. The remaining representations should also
be invariant under a $\mathbb{Z}_2$  action, which interchanges the
positive and negative chiralities. This process projects out the $Y_i$'s,
$X_{ij}^\pm$'s and $Y_{ij}^\pm$'s, while leaving the $X_i$'s and their symmetric products.
The projected Matrix algebra is non-associative but commutative
 at finite-$n$, as is the case for even
dimensional fuzzy-spheres. Here we will show that in the large-$n$ limit
non-associativity persists, unlike the case of $S^{2k}$ for which it vanishes.

We will begin by calculating the simplest
associator, $X_i * X_j * X_k$, where $*$ stands for the standard
non-associative product. This is 
\be
(X_i * X_j )* X_k - X_i * (X_j * X_k)
\ee
with the implementation of the projection performed every time a
product is calculated. The first matrix product gives
\be
X_i \cdot X_j = \mathcal P_{\mathcal R_n^+}\left[ \sum_{r=s} \rho_r (\Gamma_i \Gamma_j
  P_+)+ \sum_{r\neq s}\rho_r (\Gamma_iP_-)\rho_{s}(\Gamma_j
  P_+)\right]  \mathcal P_{\mathcal R_n^+}~~~~+~~~~ ( + \leftrightarrow - )
\ee
and after the projection
\bea\label{onetwo}
\nn X_i * X_j &=&  \mathcal P_{\mathcal R_n^+} \sum_{r=s} \rho_r (\delta_{ij}
  P_+) \mathcal  P_{\mathcal R_n^+}
 + \frac{1}{2} \mathcal P_{\mathcal R_n^+} \sum_{r\neq s} \rho_r
 (\Gamma_i
  P_- ) \rho_s(\Gamma_j P_+) \mathcal  P_{\mathcal R_n^+}\\
&& + \frac{1}{2} \mathcal P_{\mathcal R_n^+} \sum_{r\neq s} \rho_r (\Gamma_j
  P_-) \rho_s(\Gamma_i P_+) \mathcal  P_{\mathcal R_n^+} ~~~~+~~~~ ( + \leftrightarrow - )
\eea 
We  then proceed to take the ordinary product with $X_k$. 

Consider the 1-coincidence terms, where the $\Gamma_k$ acts on the same 
tensor factor as $\Gamma_i$ or $\Gamma_j$. The symmetric part of the 
product of $\Gamma$'s is clearly kept in the projected product 
defined group theoretically above. The antisymmetric part has a traceless 
part which transforms according to the Young diagram of row lengths 
$(2,1)$. The trace part transforms in $(1,0)$ and has to be kept. 
 The decomposition into traceless and trace for a 3-index tensor
 antisymmetric in two indices  is 
\bea
 A_{i[jk]} &=&  \left( A_{i[jk]} - \frac{1}{3} \delta_{ij}A_{l[lk]} - \frac{1}{3}
\delta_{ik}A_{l[jl]}  \right) + \left(  \frac{1}{3} \delta_{ij}A_{l[lk]}  + \frac{1}{3}
\delta_{ik}A_{l[jl]}  \right) \nn \\
    &=&   A_{i[jk]}^{\prime}  +
  \left( \frac{1}{3} \delta_{ij}A_{l[lk]}  + \frac{1}{3} \delta_{ik}A_{l[jl]}  \right) 
\eea
with the normalisation fixed by taking  extra contractions of the
above.  Using this, we obtain from the 1-coincidence terms\footnote{We thank 
Neil Copland  (see also \cite{bcop}) for a discussion 
which helped fix an error in the corresponding formula
 appearing  in the first version of this paper.}  
\bea\label{onecoin}  
 { ( 2n+1 ) \over 3 } \delta_{ij} X_k + { ( n-1 ) \over 6 } ( \delta_{jk} X_i + \delta_{ik} X_j ) 
\eea 
The terms with no coincidences, where the $ \Gamma_i, \Gamma_j, \Gamma_k$ 
all act in different tensor factors,  can be 
decomposed as
\be
 A_{(ij) ; k} = \frac{1}{3}\left(A_{(ij);k}  + A_{(ik); j }+ A_{(jk); i }
 \right)
 + \frac{1}{3}\left(2A_{(ij); k }-A_{(ik); j }-A_{(jk);i} \right)
\ee
It can be verified, by applying the Young Symmetriser, that the first
term corresponds to a symmetric Young diagram, while the
second to a mixed symmetry one. The traceless part of
the tensor $A_{(ij);k}$ in four dimensions  can be evaluated to be
\be
A_{(ij);k} - \frac{2\delta_{ij}}{9} A_{(ll);k}-
\frac{\delta_{ik}}{9}A_{(lj);l} - \frac{\delta_{jk}}{9}A_{(il);l}
\ee
Keeping the mixed symmetry  trace part from the non-coincident terms,  
obtained  when (\ref{onetwo})  multiplies  $X_k$ from
the left,  gives additional contributions\footnote{These contributions 
were missed in the first version of this paper.}.
 Adding these to (\ref{onecoin}) we get 
\be
(X_i * X_j )* X_k = \frac{(n^2+10n+7)}{18}\delta_{ij}X_k -
\frac{(n^2-8n+7)}{36}\left(\delta_{jk}X_i +\delta_{ik}X_j\right)  + S_{ijk}
\ee
where $S_{ijk}$ is the explicitly symmetrised product with no coincidences
\be
S_{ijk}  = \mathcal P_{\mathcal R_n^-} \sum_{r\neq s\neq t} \rho_r (\Gamma_{(i}
  P_+) \rho_s(\Gamma_j P_-)\rho_t(\Gamma_{k)} P_+) \mathcal
  P_{\mathcal R_n^+} ~~~~+~~~~ ( + \leftrightarrow - )
\ee
Similarly we find that
\be
 X_i *( X_j * X_k) =  \frac{(n^2+10n+7)}{18}\delta_{jk}X_i -
\frac{(n^2-8n+7)}{36}(\delta_{ij}X_k +
\delta_{ik}X_j)  + S_{ijk}
\ee
The difference is 
\be
(X_i * X_j )* X_k - X_i * (X_j * X_k) = \frac{n^2+4n+7}{12}(\delta_{ij}X_k-\delta_{jk}X_i)
\ee
The $X$'s should be renormalised in order to correspond to the
classical sphere co-ordinates in the large-$n$ limit. Since we have
$X_i^2\sim n^2/2$ for large-$n$ and any $D$ from (\ref{squaresD}), we
define the normalised matrices as
\be
Z_i=\frac{\sqrt{2}}{n} X_i
\ee
which gives $Z_i^2=1$. In this normalisation the associator becomes 
 \be
(Z_i * Z_j )* Z_k - Z_i * (Z_j * Z_k) =\frac{1}{6}
 \left(1+\frac{4}{n}+\frac{7}{n^2}\right)
(\delta_{ij}Z_k-\delta_{jk}Z_i)
\ee
and is obviously non-vanishing in the large-$n$ limit.

 More generally one can consider multiplying 
 $ ( S_{i_1\ldots i_p} * S_{j_1 \ldots j_q} ) * S_{k_1 \ldots k_r}$ 
 and $    S_{i_1 \ldots i_p} * ( S_{j_1 \ldots j_q}  * S_{k_1 \ldots k_r} ) $. 
It is clear from the above
discussion that the only terms of order one  (after the normalisation) 
that can appear in the
associator are the ones coming from terms with no coincidences.
 These products will, in the large-$n$ limit,  
include terms which match the classical product on the space of functions 
 on the sphere. But, as illustrated here, they will also 
 include additional terms responsible for non-associativity 
even in the large-$n$ limit. The $*$-product (discussed above) 
on the projected space of Matrices transforming as 
symmetric representations is the most obvious one available: 
the matrix product followed by projection. We have shown that it does not 
 become associative in the large-$n$ limit. There is, however, 
 another way to modify the matrix product which does become associative 
 in the large-$n$ limit. This involves keeping only the completely 
 symmetric (in $i,j,k$  etc.) part from the 
 completely non-coincident terms and is a mild modification of the 
 $*$-product discussed above. One can imagine yet other modifications.  
An alternative method for defining a non-associative product for the 
fuzzy odd-sphere, which approaches the associative one in the large-$n$ limit,
would be to start with the even sphere case for general even dimensions 
$D$  (where the prescription of matrix product followed by multiplication 
 does give vanishing non-associativity at large $n$) and then continue in $D$. 
 Whether the latter product 
is related to the alternative product contemplated above is another 
question we will leave unanswered. The eventual usefulness of such
products would be proven if we could use them in an appropriate way to 
illuminate dualities between the zero-brane and higher 
brane constructions. In the simplest 
the $D0$-$D2$ system, the non-commutative matrix 
product plays a role in the comparison of $D0$ and $D2$ actions (see for example 
\cite{prt}).

\section{Solutions to reduced action and solutions to DBI } 

In the Matrix Theory limit, in section 7, we saw that solving the 
Matrix equations of motion with the fuzzy odd-sphere 
ansatz is equivalent to solving the 
equations for the reduced action. 
We show here that the same is true for the full DBI equations of motion.
We will discuss the fuzzy-$S^3$ for concreteness, but the same
proof   applies to the case of the fuzzy-$S^5$. Consider the action arising from the expansion
of (\ref{dets3}).  The full Lagrangian will comprise of an infinite sum
of `words' $W$, consisting of products of $[\Phi, \Phi]$'s and $\partial_t
\Phi$'s, $S= - T_1 \int dt ~STr \left(\sum W \right)$. Therefore, if 
\be\label{one}
\frac{\partial W(\Phi_i=\hat R X_i)}{\partial \hat R} =  X_l ~
\frac{\partial W}{\partial \Phi_l}|_{\Phi_l= \hat R X_l}
\ee
and
\be\label{two}
\partial_t \left(\frac{\partial W(\Phi_i=\hat R X_i)}{\partial \dot{\hat
      R}}\right) = X_l ~\partial_t \left(
\frac{\partial W}{\partial(\partial_t \Phi_l)}\right)|_{\Phi_l= \hat R X_l}
\ee
then 
\be
\left[\frac{\partial}{\partial \hat R}- \partial_t \left(\frac{\partial}{\partial
  \dot{\hat R}}\right)  \right] \sum W(\Phi_i = \hat R
X_i) = X_l ~ \left[\left(\frac{\partial}{\partial \Phi_l}\right)|_{\Phi_l = \hat R
X_l}- \partial_t \left(\frac{\partial}{\partial(\partial_t \Phi_l)}
 \right)|_{\Phi_l = \hat R X_l} \right] \sum W
\ee
and a solution to the equations for the reduced action would also be a
solution to the equations coming from the full Matrix action. 

We will proceed by proving (\ref{one}) and (\ref{two}). Take a word
consisting only of $[\Phi,\Phi]$'s and expand all the commutators. The
result will be $W = (\Phi_ {A_1}\ldots\Phi_{A_m})$,
where $m$ is even and set equal to the contracted pairs of indices. Then define
\be\label{dterm}
\frac{\partial W}{\partial \Phi_l}|_{\Phi= \hat R X} = \sum_{k=1}^m \hat
D^l_{(1)}\left(\hat C_k W\right)|_{\Phi= \hat R X}
\ee
where the operator $\hat C_k$  uses the cyclic property of the
trace to rotate the $k$-th element, that is to be differentiated, to the
first slot.  $\hat D^l_{(1)}$ takes the derivative of the first term
in the word with respect to $\Phi_l$, then
sets the index of its contracted  partner equal to $l$. We have
shown that any composite operator of $SO(4)$ and $SO(6)$ with one free
index $i$ will be proportional to $X_i$. Therefore we will have that
\be
\frac{\partial W}{\partial \Phi_l}|_{\Phi= \hat R X} = \hat R^{m-1} \sum_{k=1}^m \alpha_k(W) X_l
\ee
where $\alpha_k(W)$ is some constant factor, which in general depends on the word
$W$ and $k$. One can
see that $\alpha_k(W)=\alpha(W)$, is actually independent of 
$k$. If we multiply the contribution of the $k$-th term by $ X_l$
from the left
\be
X_l  ~  \left[ \hat
D^l_{(1)}\left(\hat C_k W\right)\right]|_{\Phi= \hat R X} = \hat R^{m-1}
\alpha_k(W) X_l X_l 
\ee
On the LHS we will now have again a Casimir of $SO(D)$
\be
\hat R^{m-1} ~X_l\left[ \hat
D^l_{(1)}\left(\hat C_k W(\Phi\rightarrow X)\right)\right]= C~\hat R^{m-1} \alpha_k(W)
\ee
 As such it will obey the cyclicity property can be rotated back to
 form the original word with $\Phi\rightarrow X$
\bea
\alpha_k(W) &=& \frac{1}{C}~X_l~ \left[D^l_{(1)}\left(\hat C_k W(\Phi\rightarrow
X)\right)\right]\nn\\
&=& \frac{1}{C}W(\Phi\rightarrow X)\\
&=& \frac{1}{C} (\alpha(W)~ C)
\eea
As a consequence, every contribution in the sum (\ref{dterm}) is going
to be the same and 
\be
X_l ~\frac{\partial W}{\partial \Phi_l}|_{\Phi= \hat R X}=  m \hat R^{m-1}\alpha(W) C
\ee
It is much easier to evaluate the LHS of (\ref{one}) to get
\be
\frac{\partial W(\Phi_i=\hat R X_i)}{\partial \hat R} = m~C~\alpha(W)~\hat R^{m-1}
\ee
Exactly the same procedure can be applied to words containing time
derivatives, where $m$ is now the number of $\Phi$'s coming just from
commutators and therefore (\ref{one}) holds. Similar steps can be
carried out for the words with $m$ $\partial_t \Phi$ terms and $n$
$\Phi$ terms coming from  the expansion of commutators. We will
have
\be
\partial_t \left(
\frac{\partial W}{\partial(\partial_t \Phi_l)}\right)|_{\Phi_l= \hat R
  X_l}=  \partial_t\left(\sum_{k=1}^m \hat
S^l_{(1)}\left(\hat C_k W\right)|_{\Phi= \hat R X}\right)
\ee
where $\hat S_{(1)}^l$ takes the derivative of the first term
in the word with respect to $\partial_t \Phi_l$, then
sets the index of its contracted partner equal to $l$. This will
become
\be
 \partial_t\left(\sum_{k=1}^m \hat
S^l_{(1)}\left(\hat C_k W\right)|_{\Phi= \hat R X}\right)= 
 \left( m \hat R^{n}\dot{\hat R}^{m-1} \alpha(W) X_l\right)\dot{\phantom{P}}
\ee
and, when multiplied by $X_l$, will result into what one would get from
evaluation of the LHS of (\ref{two}), namely
\be
X_l~\partial_t \left(
\frac{\partial W}{\partial(\partial_t \Phi_l)}\right)|_{\Phi_l= \hat R
  X_l} =
 C~\alpha(W)~m\left( \dot{\hat R }^{m-1} \hat R^n\right)\dot{\phantom R}
\ee
This completes the proof that any solutions to the reduced DBI
equations of motion for $S^3$ and $S^5$ will also be solutions to the 
full Matrix equations of motion, for any $N$.
\end{appendix}

\end{document}